\documentclass[final,1p,times]{elsarticle} 
\usepackage{graphicx} 
\usepackage{amssymb} 
\usepackage{amsthm} 
\usepackage{lineno} 

\journal{Nuclear Physics A} 
\begin{document} 

\begin{frontmatter} 


\title{Aspects of thermal strange quark production: \\
the deconfinement and chiral phase transitions}

\author{Hung-Ming Tsai and Berndt M\"uller}

\address{Department of Physics, Duke University, Durham, NC 27708, USA}

\begin{abstract} 
We study the gluonic sector of the three-flavor PNJL model by
obtaining the adjoint Polyakov loop and the gluon distribution
function in the mean-field approximation. Besides, we explore the
thermal strange quark pair-production processes, $q\bar {q} \to
s\bar {s}$ and $gg \to s \bar {s}$, with the aid of the three-flavor
PNJL model. The results help us identify the temperature where the
gluonic contribution to the production rate becomes dominant, which
is an innovative phenomenon compared with the result obtained in
free perturbation theory.
\end{abstract} 

\end{frontmatter} 



\section{Introduction}

The enhanced production of the hadrons consisting of strange quarks,
especially hyperons, was proposed to be one of the signatures of the
formation quark-gluon plasma (QGP) in relativistic heavy-ion
collisions \cite{Hagedorn:1980kb}. In particular, the enhanced
pair-production of thermal strange quarks was proposed to occur as
an effect of quark and gluon deconfinement
\cite{Rafelski:1982pu,Biro:1982ud,Koch:1986ud,Matsui:1985eu}. The
predicted enhancement in the production of hyperons has been
observed in many experiments, e.~g. in
Ref.~\cite{Andersen:1999ym,Abelev:2007xp}. In temperature domains
around $200$~MeV, deconfinement and and chiral symmetry restoration
are the significant ingredients of the phase transformations of the
QCD matter, and consequently they should play important roles in the
thermal strange quark production. To elucidate their roles in the
strange quark production, we study the probability of the production
based on the thermodynamics. If the production rate is fast enough,
the probability of thermal strange quark pair-production is
$\Gamma_s=\exp (- 2 m_s/ T  )$, where $m_s$ is the constituent
strange quark mass and $T$ denotes the temperature. The
probabilities of the strange quark pair-production in the hadronic
gas (e.g. $m_s = 528$~MeV) and in the quark-gluon plasma (e.g. $m_s
= 135.7$~MeV) are plotted respectively in Fig.~\ref{figure1}(a). In
Fig.~\ref{figure1}(a), one can argue that the temperature dependence
of $\Gamma_s$ follows the dot-dashed (dashed) curve when $T<T_c$
($T>T_c$), where $T_c$ is the critical temperature of the phase
transformation. As the temperature increases, one would expect
either a sudden or smooth jump of $\Gamma_s$ in the vicinity of
$T_c$. Instead of the above argument, we focus on a more rigorous
analysis on the temperature dependence of $\Gamma_s$ in the
following contexts.

The important physics aspects of $\Gamma_s$ is sensitive not only to
the deconfinement of quarks and gluons, but also to the chiral
symmetry restoration, as can be seen from the dependence of
$\Gamma_s$ on $m_s$. The recently developed Nambu-Jona-Lasinio model
with the Polyakov loop (the PNJL model) incorporates both of the
dynamics of deconfinement and chiral symmetry restoration in
describing the phase transformations of the QCD matter
\cite{Fukushima:2003fw, Ratti:2005jh}. Therefore, in this paper, we
study the temperature dependence of $m_s$ by the three-flavor PNJL
model in the mean-field approximation, thereby obtaining $\Gamma_s$
valid for the whole temperature range. Besides, we investigate the
temperature dependence of the adjoint Polyakov loop and its
implication for the gluon distribution function. Finally, we obtain
the temperature dependence of the pair-production rates of thermal
strange quarks using the Polyakov loop-suppressed quark and gluon
distribution functions, and we identify the temperature where the
gluonic contribution to the rate becomes dominant.

\section{PNJL model} \label{sec:PNJL}

\begin{figure}[ht]
\centering

\includegraphics[width=0.45\textwidth]{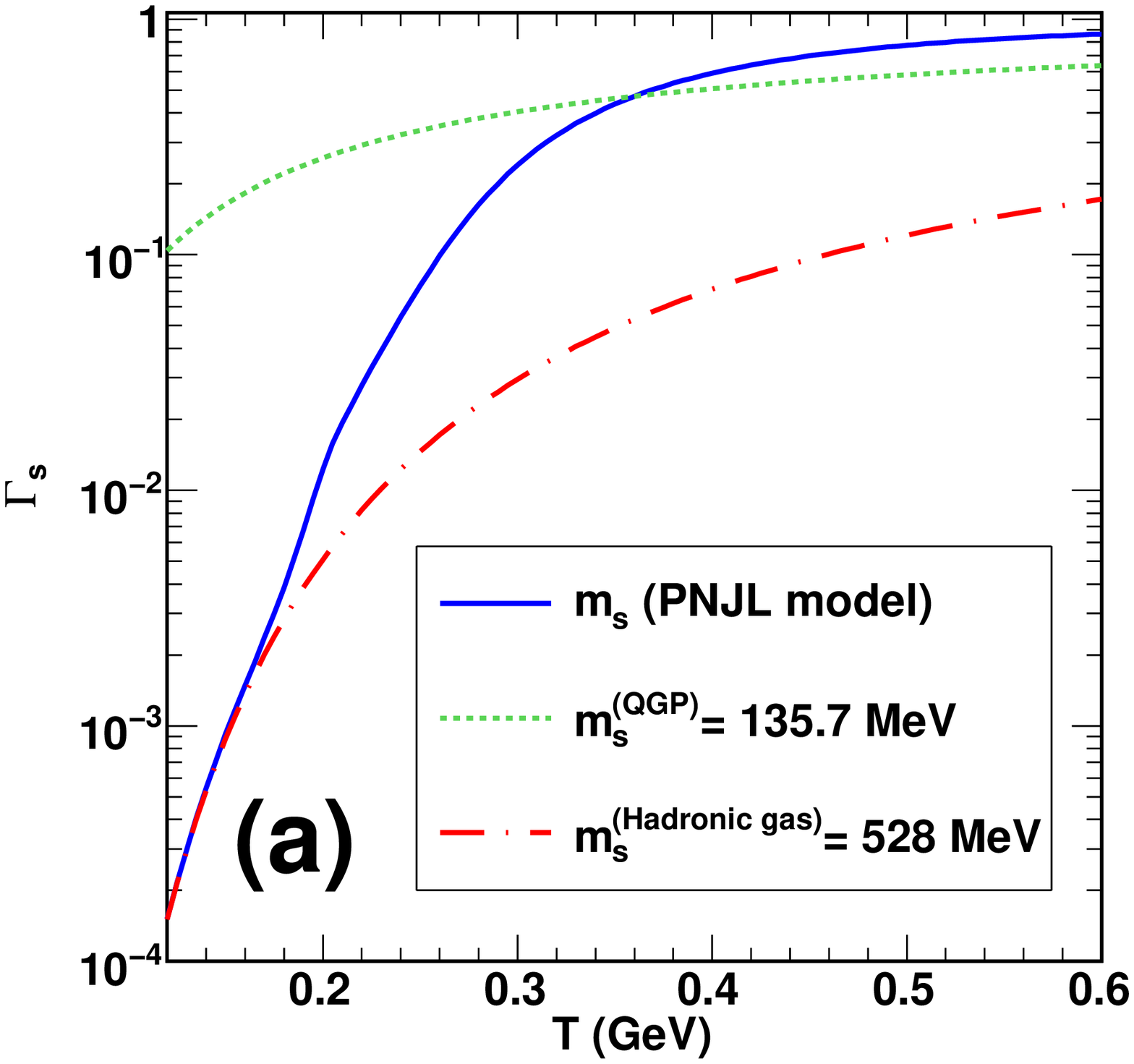}
\includegraphics[width=0.45\textwidth]{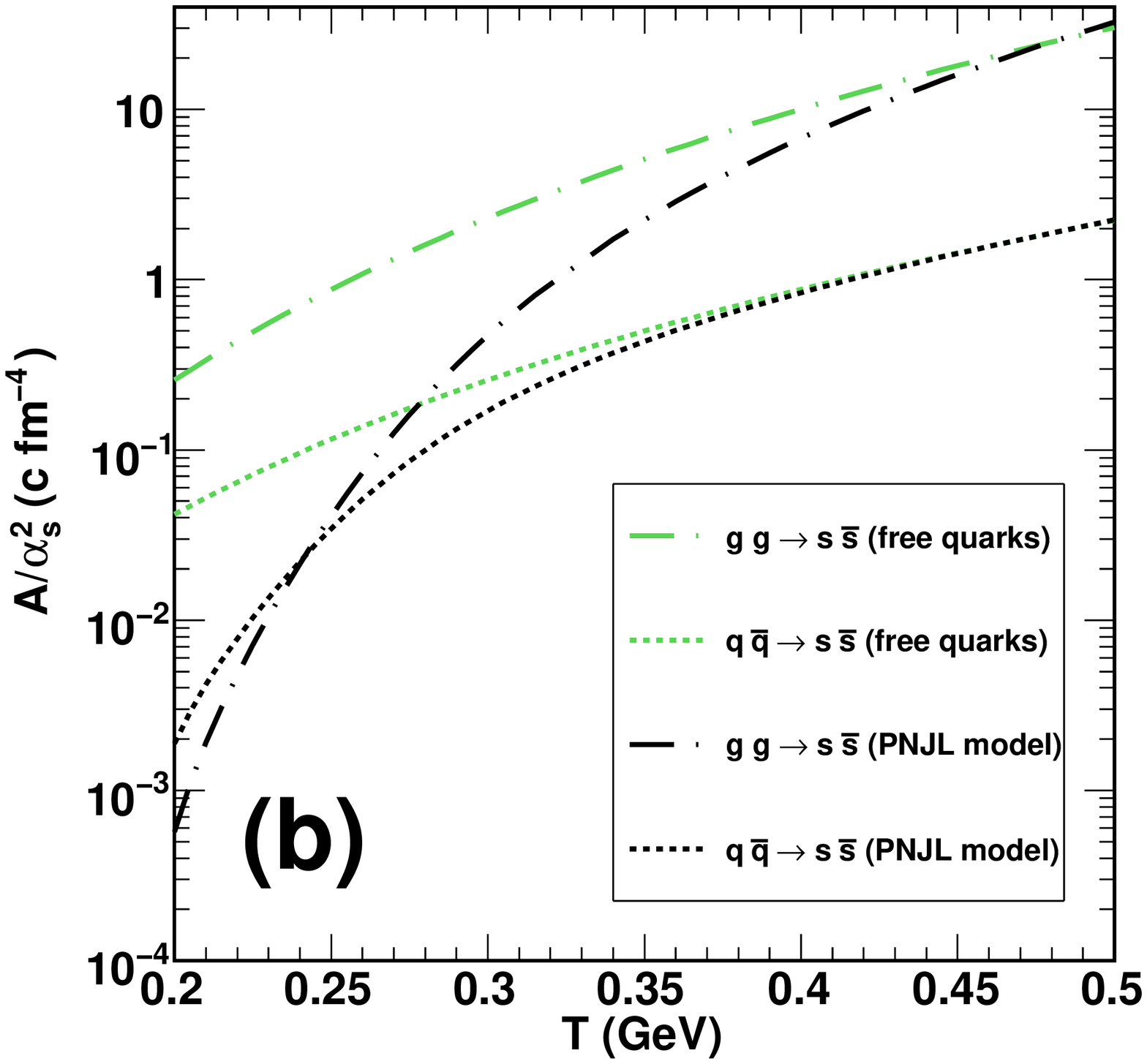}

\caption[]{(a) The probabilities of thermal strange quark
pair-production $\Gamma_s$ as functions of the temperature, with the
constituent quark mass $m_s$ obtained for the PNJL model, the
quark-gluon plasma and the hadronic gas respectively. (b) Thermal
strange quark pair-production rates divided by $\alpha_s^2$ as
functions of the temperature. The chemical potential for $u$ and $d$
quarks is $\mu=0.1$~GeV.} \label{figure1}
\end{figure}

The three-flavor PNJL model \cite{Ciminale:2007sr, Fukushima:2008wg}
incorporates the phase transformations of deconfinement and chiral
symmetry restoration in one theoretical framework. The order
parameters of deconfinement and chiral phase transitions are the
Polyakov loops and the chiral quark condensates respectively. The
Polyakov loop in color-SU(3) representation $r$ is defined as
\begin{eqnarray}
L_r = \mathcal{P} \exp \left( {ig\int_0^{1/T} {d\tau A_4 \left(
{{\mathrm {\bf x}},\tau } \right)} } \right),\label{eq:P-loop-def}
\end{eqnarray}
\noindent where $\mathcal{P}$ denotes that the exponential is
path-ordered, $T$ denotes the temperature, and $A_4 \left( {{\mathrm
{\bf x}},\tau } \right)$ is the temporal component of the
$\mathrm{SU}(3)$ gauge field in representation $r$. In particular,
$L_3 $ and $L_8 $ denote the Polyakov loops in the fundamental and
adjoint representations, respectively. The traces of the Polyakov
loops are denoted as
\begin{eqnarray}
\ell _3 = N_c^{-1} \mathrm{tr}_F L_3 , \; \; \; \; \;    \bar {\ell
}_3 = N_c^{-1} \mathrm{tr}_F L_3^\dag ,  \; \; \;\; \;     \ell _8 =
(N_c^2 - 1)^{-1} \mathrm{tr}_A L_8, \label{eq:ell_8-def}
\end{eqnarray}
where $\mathrm{tr}_F $ and $\mathrm{tr}_A $ denote the color traces
in the fundamental and adjoint representation, respectively. The
Lagrangian of the three-flavor PNJL model is \cite{Fukushima:2008wg,
Tsai:2008je}
\begin{eqnarray}
\mathcal{L} &=& \bar {\psi }\left( {i\gamma \cdot D - \hat {m}_0 }
\right)\psi - \mathcal{U} ( \ell_3, \bar{\ell}_3 ; T ) \nonumber \\
&& + \frac{g_S }{2} \sum_{a=0}^{8} \left[ {\left( {\bar {\psi
}\lambda ^a\psi } \right)^2 + \left( {\bar {\psi }i\gamma _5 \lambda
^a\psi } \right)^2} \right]  + g_D \left[ {\det \bar {\psi }\left(
{1 - \gamma _5 } \right)\psi + h.c.} \right], \label{eq:PNJL-Lagr}
\end{eqnarray}
\noindent where $D_\mu = \partial _\mu - g\delta _{\mu 4} A_4$ is
the gauge-covariant derivative, $\mathcal{U} ( \ell_3, \bar{\ell}_3
; T )$ is the effective potential for the Polyakov loop, and $\hat
{m}_0 = \mathrm{diag}(m_{u,0} ,m_{d,0} ,m_{s,0} )$. $g_S$ and $g_D$
are shown explicitly in Ref.~\cite{Fukushima:2008wg, Tsai:2008je}.
In the limit of isospin symmetry, $m_{u,0} = m_{d,0} = m_{q,0} $.
From the above Lagrangian, we obtain the grand canonical
thermodynamic potential per unit volume, $\Omega$, in the mean field
approximation. We obtain the temperature dependence of the order
parameters, $\langle {\bar {q}q} \rangle$, $\langle {\bar {s}s}
\rangle$, $\langle {\ell _3 } \rangle $ and $\langle {\bar {\ell}_3
} \rangle $, by minimizing $\Omega$ with respect to these order
parameters and solving the equations simultaneously
\cite{Tsai:2008je}. Besides, the constituent quark masses can be
expressed as functions of the current masses and the chiral
condensates
\begin{eqnarray}
m_q &=& m_{q,0} - 2g_S \langle {\bar {q}q} \rangle - 2g_D \langle
{\bar {q}q} \rangle \langle {\bar {s}s} \rangle, \qquad (q=u,d)
\label{eq:mq}
\\
m_s &=& m_{s,0} - 2g_S \langle {\bar {s}s} \rangle - 2g_D \langle
{\bar {q}q} \rangle ^2,\label{eq:ms}
\end{eqnarray}
where  $\langle {\bar {u}u} \rangle = \langle {\bar {d}d} \rangle $,
$m_{u,0} = m_{d,0}$, $m_q^0 = 0.0055$~GeV and $m_s^0 = 0.1357$~GeV
\cite{Fukushima:2008wg}. The temperature dependence of $\langle
{\bar {q}q} \rangle$, $\langle {\bar {s}s} \rangle$, $\langle {\ell
_3 } \rangle $, $\langle {\bar {\ell}_3 } \rangle $, $m_q$ and $m_s$
are depicted in Ref.~\cite{Tsai:2008je}. In the PNJL model, we
obtain $\Gamma_s$ as a function of temperature with $m_s$ input by
Eq.~(\ref{eq:ms}), as shown in Fig.~\ref{figure1}. We note that
$\Gamma_s$ in the PNJL model exceeds $\Gamma_s$ of deconfined quarks
for $T > 360$~MeV, which is due to an artifact of the PNJL model
that $m_q < m_{q,0}$ and $m_s < m_{s,0}$ for $T > 360$~MeV.
Moreover, in the PNJL model, $\Gamma_s \to 1$ at high temperature.

\section{Adjoint Polyakov loop and gluon distribution function}
We obtain the thermal average of the adjoint Polyakov loop from that
of the fundamental Polyakov loop by the following self-consistent
procedure \cite{Tsai:2008je}. First, we define the thermal average
with respect to the weight function, $ W( \ell _3 ;T) = \exp \left(
{6 \, d \, \beta _3 \, \langle {\ell _3 } \rangle \,
\mathrm{Re}(\ell _3)} \right),$ where $d = 3$ and $\beta_3 ( T )$ is
a fit parameter depending on temperature \cite{Gocksch:1984yk,
Gupta:2007ax}. Then, the thermal average of the fundamental Polyakov
loop gives the constraint equation, $ \langle \ell _3 \rangle  =
\langle  \mathrm{tr}_F L_3 / 3 \rangle _W $, where $\langle \ell _3
\rangle$ is, at each temperature, input by its mean-field value
obtained in Sec.~\ref{sec:PNJL}. We obtain $\beta_3 (T)$ by solving
this constraint equation, and then we evaluate $\langle \ell_8
\rangle$ as a function of the temperature. We verify that, with this
averaging procedure, the adjoint and fundamental Polyakov loops
satisfy the Casimir scaling, $\left\langle {{\ell _8}} \right\rangle
= {\left\langle {{\ell _3}} \right\rangle ^{9/4}}$, which is
observed in lattice QCD \cite{Gupta:2007ax, Gupta:2006qm}. Moreover,
we obtain the color-averaged gluon distribution function
\cite{Tsai:2008je}
\begin{equation}
f_g \left( k \right) = \frac{1}{8}\sum\limits_{n = 1}^\infty
{\langle {\mathrm{tr}_A L_8^n } \rangle \exp \left( { - {n\left|
{\mathrm {\bf k}} \right| / T}} \right)}. \label{eq:gluon-dist}
\end{equation}
We show that the color-averaged quark and gluon distribution
functions are both suppressed by the Polyakov loops, and the degree
of suppression for the gluon distribution is higher than that of the
quark distribution \cite{Tsai:2008je}.

\section{Thermal strange quark pair-production}

The pair-production of thermal strange quarks is contributed by two
processes, $q\bar {q} \to s \bar {s}$ and $gg \to s \bar {s}$. The
production rates was first obtained by QCD perturbation theory for
free quarks \cite{Rafelski:1982pu}. In this paper, we take into
account the effects of deconfinement and chiral symmetry restoration
on the production rates. We obtain the strange quark production rate
per unit volume by
\begin{eqnarray}
A= \frac{dN}{ dtd^3x} = A_q + A_g,
\end{eqnarray}
\noindent where the explicit expressions for $A_q$ and $A_g$ are
shown in Ref.~\cite{Tsai:2008je}. In Fig.~\ref{figure1}(b), we plot
the temperature dependence of the production rates divided by
$\alpha_s^2$ for the PNJL model and those for free quarks
respectively. The production rates for the PNJL model are smaller
(larger) than those for free quarks when $T<480$~MeV ($T>480$~MeV).
This phenomenon is due to either the addition or the competition of
the following two effects. One effect is the suppression of the
thermal quark and gluon excitations by the Polyakov loops, and the
other effect is that, in the PNJL model, $m_s>m_{s,0}$
($m_s<m_{s,0}$) when $T<360$~MeV ($T>360$~MeV). The later effect
surpasses the former one when $T>480$~MeV so that the production
rates for the PNJL model become larger than those for free quarks.
Besides, as proposed in \cite{Rafelski:1982pu,Koch:1986ud}, the
production rates for free quarks are enhanced in the deconfined
phase, and the production rates for $gg \to s\bar {s}$ is dominant
at all temperatures. In the PNJL model, the enhanced production is
also obtained, but the production rates for $q\bar {q} \to s\bar
{s}$ and $gg \to s\bar {s}$ cross over at $T_r \approx 240$~MeV. The
production rate for $q\bar {q} \to s\bar {s}$ ($gg \to s\bar {s}$)
is dominant when $T<T_r$ ($T>T_r$). Because the threshold $T_r$ is
well beyond the temperature range in which the chiral phase
transition occurs, and our neglect of the contribution from
collective (hadronic) modes in the transition region near $T_c$
appears justified. Moreover, when $T<T_c$, the production rates for
the PNJL model are very small because quark and gluon quasiparticles
are strongly suppressed below $T_c$. In this temperature region,
strangeness production is dominated by hadronic reactions, which
were investigated by Rehberg {\em et al.} in the NJL model
\cite{Rehberg:1995kh}.


\section*{Acknowledgments} 
This work was supported in part by the U.~S.~Department of Energy
under grant DE-FG02-05ER41367. We thank Kenji Fukushima for several
enlightening discussions about the PNJL model. One of us (HMT)
acknowledges the support from the organizers of the Quark Matter
2009 conference.

\end{document}